\documentclass[final,5p,times,twocolumn,numbers]{elsarticle}

\usepackage{amssymb}
\usepackage{lipsum}
\usepackage{amsmath}
\usepackage{graphicx}
\usepackage{caption}
\usepackage{subcaption}
\usepackage{float}
\usepackage{color}
\usepackage{booktabs}

\journal{Nuclear Physics B}

\begin{document}
\begin{frontmatter}
\title{Probing Near-Threshold $s$-Wave Components in Heavy Nuclei via Coulomb-Assisted Neutron Transfer}
\author[first]{Yuki Nakanishi\corref{cor1}}
\ead{yukin@rcnp.osaka-u.ac.jp}
\affiliation[first]{organization={Department of Physics, The University of Osaka},
            addressline={Machikaneyama-cho 1-1},
            city={Toyonaka},
            postcode={560-0043},
            state={Osaka},
            country={Japan}}
\cortext[cor1]{Corresponding author}
\author[first,second]{Junki Tanaka}
\affiliation[second]{organization={Research Center for Nuclear Physics, The University of Osaka},
            addressline={Mihogaoka 10-1}, 
            city={Ibaraki},
            postcode={567-0047}, 
            state={Osaka},
            country={Japan}}
\author[first,second]{Atsushi Tamii}
\author[third,forth]{Shimpei Endo}
\affiliation[third]{organization={Department of Engineering Science, University of Electro-Communications},
            addressline={Chofugaoka 1-5-1}, 
            city={Chofu},
            postcode={182-8585}, 
            state={Tokyo},
            country={Japan}}
\affiliation[forth]{organization={Institute for Advanced Science, University of Electro-Communications},
            addressline={Chofugaoka 1-5-1}, 
            city={Chofu},
            postcode={182-8585}, 
            state={Tokyo},
            country={Japan}}
\begin{abstract}
\small
We propose a method to probe weakly bound s-wave neutron components near the neutron emission threshold in heavy nuclei using Coulomb-assisted neutron transfer reactions. Weakly bound s-wave neutrons have large asymptotic amplitudes, which are difficult to access directly with conventional methods.
This work focuses on the $(d,p)$ reaction at low incident energies and backward angles, where the reaction is localized in the nuclear exterior due to the Coulomb barrier. Under these conditions, the transition amplitude becomes sensitive to the asymptotic part of the single-particle wave function.
Finite-range DWBA calculations show that the cross section for weakly bound states exhibits a weak dependence on incident energy, while that for strongly bound states decreases rapidly with decreasing energy. Contributions from orbitals with $l \geq 1$ are suppressed by the centrifugal barrier, resulting in selectivity for s-wave components.
This method provides a probe of the strength distribution of weakly bound s-wave components near threshold and the asymptotic structure of their wave functions.
\end{abstract}

\begin{keyword}
$(d,p)$ reaction \sep peripheral reaction \sep s-wave \sep neutron transfer \sep near threshold \sep asymptotic structure
\end{keyword}

\end{frontmatter}

\section{Introduction}
\label{introduction}
\begin{figure}[h!]
    \centering
    \includegraphics[width=1.0\linewidth]{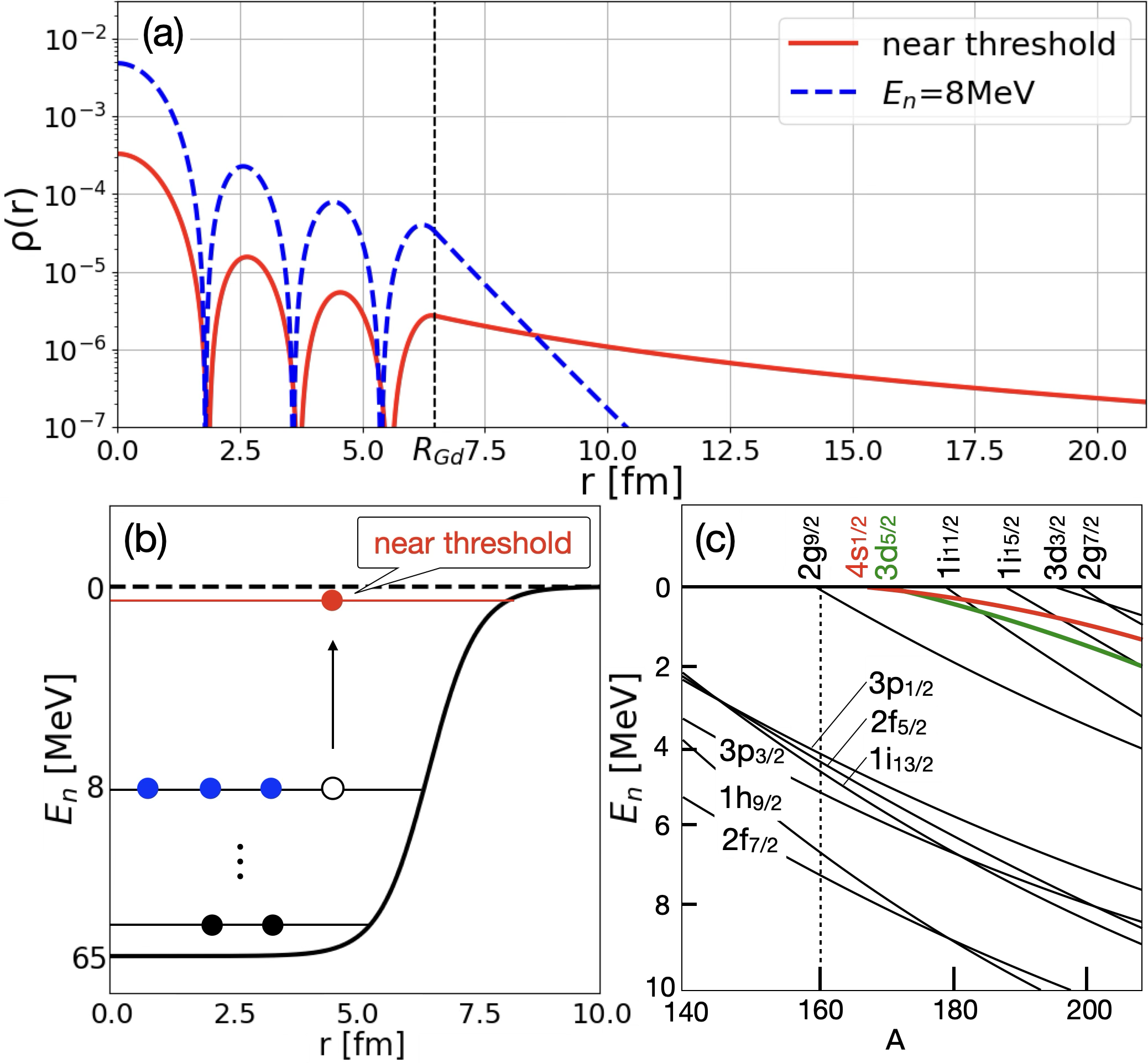}\vspace{-5pt}
    \caption{(a) Density distributions for strongly and weakly bound cases for a three-node 4$s_{1/2}$ orbital. The weakly bound case exhibits a pronounced spatial expansion beyond the nuclear radius $R_{\mathrm{Gd}}$. (b) Schematic illustration of a weakly bound neutron component near the threshold induced by a 1$p$--1$h$ excitation in a Woods--Saxon potential. (c) Mass dependence of single-particle energies showing that the 4$s_{1/2}$ orbital (red line) approaches the neutron separation threshold in the $A \sim$ 160 region.  The green line corresponds to the 3$d_{5/2}$ orbital ~\cite{BohrMottelson}.}
    \label{fig:wf}
\end{figure}

Near particle emission thresholds, exotic nuclear structures such as cluster and halo configurations are known to emerge~\cite{Ikeda1968,Tanihata2013}. These phenomena are closely related to the spatial extension of wave functions in weakly bound systems. For a weakly bound single-particle $s$ wave, the wave function outside the mean-field potential is expressed as
\vspace{-4pt}
\begin{equation}
\phi(r)=A\frac{e^{-\kappa r}}{r},
\quad with \quad
\kappa=\sqrt{\frac{2mE_n}{\hbar^2}}.
\end{equation}
\vspace{-4pt}
Here, $E_n$ is the neutron binding energy. The weak-binding behavior is illustrated by the density distribution $\rho(r)$ in Fig.~\ref{fig:wf}(a). Such structures are established in light neutron-rich nuclei, exemplified by the halo in $^{11}\mathrm{Li}$~\cite{Tanihata1985}. 

Similar phenomena may also arise in excited states of nuclei when particle--hole excitations populate shallow levels near the neutron separation threshold (Fig.~\ref{fig:wf}(b)). In light nuclei, $(d,p)$ reaction experiments have been used to investigate their relation to halo structures~\cite{Fulmer1964,Liu2001,Belyaeva2014}. To extend to heavy nuclei, we focus on the mass region $A \sim 160$, where systematic calculations of single-particle energies predict that the 4$s_{1/2}$ orbital lies near the neutron separation threshold~\cite{BohrMottelson,Campbell1960} (Fig.~\ref{fig:wf}(c)). In $^{155}$Gd and $^{157}$Gd, large thermal neutron capture cross sections associated with the $s$-wave component originating from the 4$s_{1/2}$ orbital have been reported, indicating that the proximity of the resonance energy to the threshold is related to the enhancement of the cross section~\cite{BohrMottelson,Mughabghab,Yamaguchi2024}. In this mass region, however, the level density is high, and the single-particle s-wave strength is expected to be fragmented over many states~\cite{Feshbach}. Although large thermal neutron capture cross sections suggest the presence of near-threshold s-wave components, they do not directly probe the spatial extension of the wave function, motivating the need for a reaction mechanism that selectively samples the nuclear exterior.

In this study, we propose a Coulomb-assisted $(d,p)$ reaction at low incident energies and backward angles as a probe of near-threshold s-wave strength. Under these conditions, the reaction is localized in the nuclear exterior, and the transition amplitude becomes sensitive to the asymptotic part of the neutron wave function, because the reaction amplitude is governed by the spatial overlap between the distorted waves and the tail of the bound-state wave function. The feasibility of this method is examined for nuclei in the $A \sim 160$ region using finite-range DWBA calculations.
Experimentally, the excitation energy of the residual nucleus is reconstructed from the measured energy and angle of backward-emitted protons. The incident-energy dependence of the backward-angle cross section for near-threshold states is then used to probe the spatial extension of the transferred neutron component.

\section{Coulomb-dominated peripheral reaction}
The term “Coulomb-assisted” refers to reaction conditions in which the Coulomb barrier suppresses the penetration of the incident deuteron into the nuclear interior. We therefore consider low incident energies and large scattering angles, under which the transfer amplitude is shifted toward the nuclear exterior and becomes sensitive to the asymptotic part of the neutron wave function. To examine this condition, differential cross sections for the $(d,p)$ reaction at various incident energies were calculated using the finite-range DWBA code \textsc{FRESCO} \cite{FRESCO}. The reaction was analyzed using the finite-range distorted-wave Born approximation (DWBA) in the post representation,
\begin{equation}
T =
\left\langle
\chi_{pB}^{(-)}(\mathbf{R}_{pB})\,\phi_{nA}(\mathbf{r}_{nA})
\right|
V_{nA}(\mathbf{r}_{nA})
\left|
\chi_{dA}^{(+)}(\mathbf{R}_{dA})\,\phi_d(\mathbf{r}_{np})
\right\rangle ,
\end{equation}
where $\chi_{dA}^{(+)}$ and $\chi_{pB}^{(-)}$ denote the distorted waves in the entrance and exit channels, respectively, $\phi_d$ is the internal wave function of the deuteron, and $\phi_{nA}$ is the neutron bound-state wave function in the residual nucleus. The transition operator $V_{nA}$ corresponds to the neutron--core binding potential used to generate the bound-state wave function $\phi_{nA}$. In the present calculation, $V_{nA}$ was described by a Woods--Saxon potential with geometry parameters $r_0 = 1.25$~fm and $a = 0.65$~fm, and the depth was adjusted to reproduce the binding energy for each excitation energy $E_n$. The optical potential parameters used in the calculations are summarized in Table~\ref{table:DWBAparameters}.

\begin{figure}[h]
    \centering
    \includegraphics[width=1.0\linewidth]{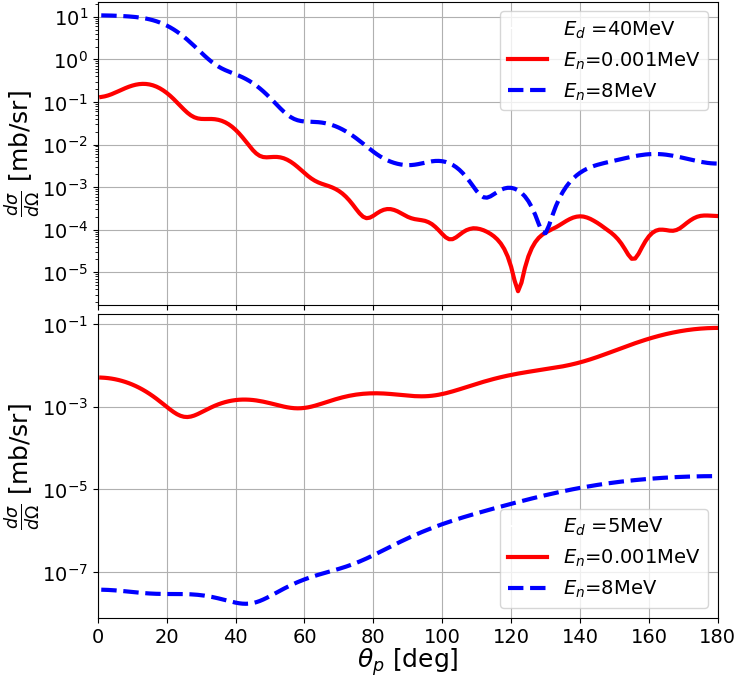}\vspace{-10pt}
    \caption{Angular distributions of differential cross sections for the $(d,p)$ reaction at different incident energies and neutron separation energies. The upper panel shows the results at $E_d=40$ MeV, while the lower panel corresponds to $E_d=5$ MeV. The solid-red and dotted-blue curves correspond to transitions to weakly and strongly bound states, respectively.}
    \label{fig:DWBA40MeV}
\end{figure}

\begin{table}[h]
\centering
\caption{Woods--Saxon optical potential parameters (real/imaginary parts) for deuterons~\cite{Daehnick1981} and protons~\cite{Bechetti1969} used in the DWBA calculations.}
\label{table:DWBAparameters}
\vspace{2pt}
\begin{tabular}{llll}
\hline
\vspace{1pt}
Channel & Depth (V/W) & Radial Para. & Diffuseness\\
\hline
$d+{}^{157}\mathrm{Gd}$ & 90.0 / 21.0 & 1.15/1.34 & 0.81/0.68 \\
$p+{}^{158}\mathrm{Gd}$ & 47.9 / 11.5 & 1.25/1.25 & 0.65/0.47 \\
\hline
\end{tabular}
\end{table}
In the present calculations, the spectroscopic factor was set to unity in order to examine the reaction selectivity independently of nuclear-structure fragmentation. The numerical parameters in FRESCO, including the matching radius and partial waves, were chosen to ensure convergence of the calculated cross sections.
The calculated results are shown in Fig.~\ref{fig:DWBA40MeV}. The red and blue curves correspond to transitions to weakly and strongly bound states, respectively.

As shown in Fig.~\ref{fig:DWBA40MeV}, at an incident energy of 40 MeV, the differential cross section for transitions to weakly bound states is smaller than that for strongly bound states over all angles. In this case, the incident wave penetrates into the nuclear interior, and the reaction is governed by overlap with wave functions localized inside the nucleus.

When the incident energy is reduced to 5~MeV, the cross section for weakly bound states exceeds that for strongly bound states and exhibits a pronounced peak at backward angles, particularly near $180^\circ$. This inversion reflects the transition from an interior-dominated reaction at high energy to a peripheral reaction at low energy, driven by the reduced penetration of the deuteron wave function due to the Coulomb barrier at low energies.

This behavior demonstrates that backward angles provide enhanced sensitivity to weakly bound components under peripheral conditions.

\section{Incident-energy dependence of the reaction cross sections}

\begin{figure}[h]
\centering
\includegraphics[width=1.0\linewidth]
{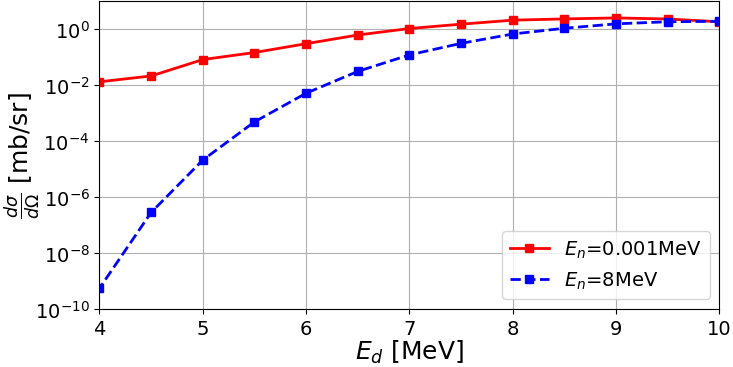}\vspace{-10pt}
\caption{Incident-energy dependence of the $(d,p)$ reaction cross section at $180^\circ$. The solid-red and dotted-blue curves correspond to transitions to weakly and strongly bound states, respectively.
}
\label{fig:180degCS}
\end{figure}

We next examine how the sensitivity to the asymptotic wave function manifests itself in the incident-energy dependence of the cross section. Figure~\ref{fig:180degCS} shows the results of finite-range DWBA calculations~\cite{FRESCO}. The vertical axis represents the differential cross section at $180^\circ$, and the horizontal axis is the incident deuteron energy. The blue curve corresponds to a strongly bound state with $E_n = 8\,\mathrm{MeV}$, while the red curve represents transitions to a weakly bound state.

For the strongly bound state, the cross section falls rapidly below $E_d$ $\sim$ 8 MeV, whereas the weakly bound s-wave transition remains sizable down to lower energies. This behavior reflects the larger asymptotic amplitude of the weakly bound s wave.
In contrast to strongly bound states, the cross section for weakly bound s-wave components varies only slowly with decreasing incident energy, providing an experimental signature of their spatially extended nature.

In realistic heavy nuclei, the single-particle s-wave strength is expected to be fragmented over many states near the neutron threshold. In the present calculation, a single-particle weakly bound s-wave orbital is used as a representative case to demonstrate the reaction sensitivity to the asymptotic wave function. Experimentally, the same sensitivity can be used to identify the distribution of spatially extended s-wave strength.

\section{Selectivity to orbital angular momentum}
We next examine whether the proposed condition is selective not only to weak binding but also to orbital angular momentum.

For orbitals with $l \geq 1$, the centrifugal barrier suppresses the spatial extension of the wave function. As a result, even at the same binding energy, the amplitude in the exterior region is smaller than that of an $s$ wave. This difference in spatial distribution is reflected in the overlap with the incident wave localized in the exterior region.

\begin{figure}[H]
\centering
\includegraphics[width=1.0\linewidth]{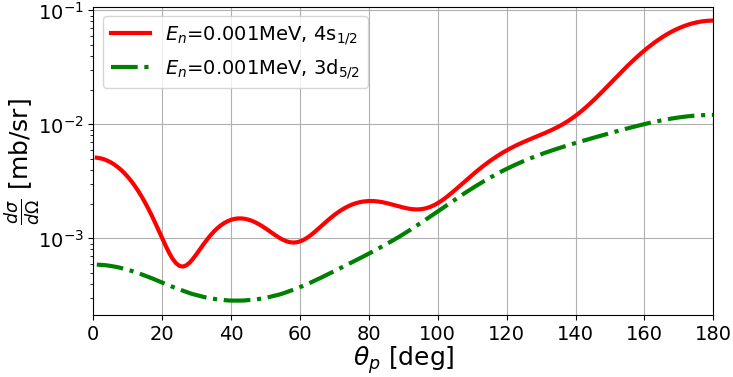}\vspace{-10pt}
\caption{
Angular distributions of differential cross sections for the $(d,p)$ reaction at $E_d\,=\,5\,\text{MeV}$. The solid-red and dash–dot green curves correspond to transitions to the $4s_{1/2}$ and $3d_{5/2}$ orbitals, respectively.
}
\label{fig:compdw}
\end{figure}

To evaluate this effect, neutron transfer to the weakly bound $4s_{1/2}$ orbital and the nearby $3d_{5/2}$ orbital (see Fig.~\ref{fig:wf}(c)) was compared in the mass region $A \sim 160$. Figure~\ref{fig:compdw} shows finite-range DWBA results at an incident energy of 5~MeV. Both orbitals are assumed to be weakly bound, and the spectroscopic factor is set to unity. The red and green curves correspond to transitions to $4s_{1/2}$ and $3d_{5/2}$, respectively.

Both cases show maxima at backward angles, particularly near $180^\circ$. In the present calculation, the cross section for the $3d_{5/2}$ orbital is reduced to about 0.15 times that for the $4s_{1/2}$ orbital. This reduction arises because the $d$-wave function decays rapidly in the exterior region due to the centrifugal barrier, leading to a smaller overlap.
Therefore, the proposed condition provides selectivity not only to weak binding but also to s-wave character.



\section{Conclusion and outlook}
We propose a Coulomb-assisted $(d,p)$ reaction as a probe of weakly bound $s$-wave components near the neutron threshold in heavy nuclei.
Finite-range DWBA calculations reveal two characteristic signatures: enhanced backward-angle cross sections at low incident energies and a significantly weaker energy dependence compared to strongly bound states. These features originate from the large asymptotic amplitude of weakly bound $s$-wave functions, together with the suppression of interior contributions by the Coulomb barrier. The additional centrifugal suppression of $l \geq 1$ components further enhances the selectivity for $s$ waves.
Although this work focuses on neutron transfer on ${}^{157}\mathrm{Gd}$, leading to states in ${}^{158}\mathrm{Gd}$, the mechanism is general and applicable to nuclei with near-threshold $s$-wave orbitals. Systematic measurements along isotopic chains could provide constraints on the distribution and asymptotic structure of $s$-wave strength near the threshold.
This provides a new observable to experimentally access the asymptotic structure of neutron wave functions in heavy nuclei.

For realistic experimental applications, several effects should be taken into account. In particular, background contributions from Coulomb-induced breakup processes may affect the measured cross sections. Furthermore, a more detailed analysis of experimental data will likely require coupled-channel treatments, such as CDCC calculations, to properly describe continuum effects and reaction dynamics.

\section*{Acknowledgements}
This work was supported by JSPS KAKENHI Grant Numbers JP25KK0050 and JP24K17069. S.E. acknowledges additional support from JSPS KAKENHI Grant Numbers JP23H01174 and JP25K00217 and from the Matsuo Foundation. The authors are grateful to Tokuro Fukui, Nori Aoi, Susumu Shimoura, Shinsuke Ota, Nobuyuki Kobayashi, Juzo Zenihiro, Yohei Matsuda, and Futoshi Minato for valuable discussions and insightful comments on this work.


\end{document}